\documentclass[pre,aps,twocolumn,superscriptaddress]{revtex4-1}
\pdfoutput=1
\usepackage{amssymb}
\usepackage{epsfig}
\usepackage{amsmath}
\usepackage{subfigure}
\usepackage{graphicx}
\usepackage{textcomp}
\usepackage{url}
\usepackage{float}
\usepackage{color}
\usepackage{cases}
\usepackage[normalem]{ulem}

\usepackage[titletoc,title]{appendix}
\usepackage[colorlinks=true, urlcolor=blue, anchorcolor=blue, citecolor=blue,filecolor=blue,linkcolor=blue,menucolor=blue
]{hyperref}

\newcommand{\be}{\begin{equation}}
\newcommand{\ee}{\end{equation}}
\newcommand{\bea}{\begin{eqnarray}}
\newcommand{\eea}{\end{eqnarray}}

\begin{document}
\title{Relaxation and fluctuations of a mass- and dipole-conserving stochastic lattice gas}
\author{Baruch Meerson}
\email{meerson@mail.huji.ac.il}
\affiliation{Racah Institute of Physics, Hebrew University of
Jerusalem, Jerusalem 91904, Israel}

\begin{abstract}

Han et al. [Phys. Rev. Lett. \textbf{132}, 137102 (2024)] have recently introduced a classical stochastic lattice gas model which, in addition to particle conservation, also conserves the particles' dipole moment. Because of its intrinsic nonlinearity this model exhibits unusual macroscopic scaling behaviors, different from those of lattice gases that conserve only the number of particles.  Here we investigate some basic relaxation and fluctuation properties of this model at large scales and at long times. These properties crucially depend on whether the total number of particles is infinite or finite. We find similarity solutions, describing relaxation of the dipole-conserving gas (DCG) in several standard settings. A major part of our effort is an extension to this model of the macroscopic fluctuation theory (MFT), previously developed for lattice gases where only the number of particles is conserved. We apply the MFT to the calculation of the variance of nonequilibrium fluctuations of the excess number of particles on the positive semi-axis when starting from an (either deterministic, or random) constant density at $t=0$. Using the MFT, we also identify the equilibrium Boltzmann-Gibbs distribution for the DCG. Finally, based on these results, we determine the probability distribution of, and the most probable density history leading to, a large deviation in the form of a macroscopic void of a given size in an initially uniform DCG at equilibrium.
\end{abstract}
\maketitle

\section{Introduction}
\label{sec:Intro}

Lattice gases, composed of particles undergoing symmetric random hopping, exhibit diffusive behavior at large scales and long times \cite{Spohn91,Anna,Kipnis99,Schutz00}. As these models conserve the number of particles locally, the gas density $\rho(\mathbf{x},t)$ at large scales and at long times obeys the continuity equation
\be
\partial_t \rho + \nabla\cdot {\mathbf J}=0
\label{continuity}
\ee
with the diffusion current
\be
{\mathbf J}= - D(\rho) \nabla \rho\,.
\label{J}
\ee
For a limited class of models -- generically for the so called gradient lattice gases \cite{Spohn91,Kipnis99,Gabrielli18} -- the diffusion coefficient $D(\rho)$ can be calculated from the microscopic model exactly. In simple cases, which include noninteracting random walkers, $D(\rho)$ is independent of the density, and Eqs.~\eqref{continuity} and~(\ref{J}) reduce to the simple (linear) diffusion equation
\begin{equation}
\label{DE}
\partial_t \rho = D \Delta \rho\,,
\end{equation}
where $\Delta$ is the Laplacian operator.

How is the simple diffusion model modified if the lattice gas is constrained to satisfy additional local conservation laws (besides the conservation of the number of particles)? Systems with additional conservation laws have attracted much interest in recent years. Many studies concerned quantum systems with infinitely many conservation laws, specifically integrable systems in one spatial dimension. The corresponding generalized hydrodynamics  is an active research area, see Refs. \cite{Doyon20c,Dubail22} for reviews. The generalized hydrodynamics was also used for the description of the hydrodynamic behavior in classical integrable systems in one spatial dimension \cite{Doyon19,Doyon20a,Spohn23}. Quantum higher-moment conserving models attracted much attention in the context of fractons \cite{Chamon05,Pretko17,Gromov19,Williamson19,Lake22,Williamson23,Gromov24};  see Refs.~\cite{Lucas20,Knap20,Lucas21,Surowka23,Morningstar23} for a sample of articles about the emerging `fracton hydrodynamics'. Lattice gases with additional conservation laws provide a convenient and flexible theoretical platform for a systematic derivation (rather than a postulation) of a `fracton hydrodynamics'.

Han et al. \cite{HLR2024} have recently introduced an interesting stochastic lattice gas model which manifestly conserves, in addition to the number of particles, their dipole moment. The microscopic model of Han et al. involves a continuous-time lattice gas, where a randomly chosen pair of neighboring particles randomly hop in opposite directions in pairs, so that their center of mass is conserved. Han et al. obtained the continuum limit of this model via a standard derivative expansion of the deterministic rate equation for the particle densities on each site. In one spatial dimension, the resulting large-scale deterministic description is given by a nonlinear partial differential equation (PDE) of fourth order \cite{HLR2024}:
\begin{eqnarray}
  \partial_t\rho &=& -D\partial_x^2\left[\rho \partial_x^2 \rho-(\partial_x \rho)^2\right] \nonumber \\
  &=& -D \partial_{x}^{2}\left(\rho^2 \partial_x^2 \ln \rho\right) \nonumber \\
   &=& - D \left[\rho \partial_x^4 \rho-(\partial_x^2 \rho)^2\right]\,. \label{MF}
\end{eqnarray}
As one can see, here the simple first-order continuity equation (\ref{continuity}) gives way to the second-order equation \cite{Pretko17}
\begin{equation}
\label{D:continuity}
\partial_t \rho +  \partial_x^2 J = 0\,,
\end{equation}
with the current
\begin{equation}
\label{D:J}
 J = D\rho^2\partial_x^2 \ln \rho\,.
\end{equation}
Equation~(\ref{MF}) can be generalized to an arbitrary spatial dimension \cite{HLR2024}:
\begin{equation}
\label{Dipole-d}
\partial_t \rho =- D [\rho\Delta^2 \rho-(\Delta \rho)^2]\,,
\end{equation}
where $\Delta \equiv \nabla^2$ is the Laplace's operator.  The transport coefficient $D$ comes from the microscopic model, and it has the units of $\text{length}^{d+4}/\text{time}$, where $d$ is the dimension of space \cite{jamming}.

Because of an interplay of the nonlinearity and the fourth-order spatial derivative, Eqs.~(\ref{MF}) and (\ref{Dipole-d}) exhibit unusual macroscopic scaling behaviors \cite{HLR2024}. Here we will further explore these scaling behaviors
by deriving some similarity solutions to these equations, which describe relaxation of the dipole-conserving gas (DCG) in several standard settings. As we will see, the scaling properties of the relaxation dynamics crucially depend on whether the total number of particles is infinite or finite.

The main focus of this work, however, is on large-scale fluctuations in the DCG. A study of fluctuations obviously requires going beyond the deterministic limit, described by Eqs.~(\ref{MF})  and (\ref{Dipole-d}). Han et al. \cite{HLR2024} have already made this important step by deriving, from a microscopic lattice gas model, a Langevin equation for this system, see Eq.~(\ref{Langevin0}) below. In addition to the terms present in the deterministic equation~(\ref{MF}), this  stochastic PDE also includes a noise term. A similar-in-spirit Langevin description of the mass-only conserving lattice gases is known  by the name of `fluctuational hydrodynamics' \cite{Spohn91,Anna,Kipnis99,Schutz00}. Starting from the Langevin equation~(\ref{Langevin0}), here we develop a macroscopic fluctuation theory (MFT), which is suitable for studying large deviations of different fluctuating quantities in the DCG. In the mass-only conserving lattice gases the corresponding MFT was developed by Jona-Lasinio et al., see Ref. \cite{bertini} for a review, and it has been employed and further developed in numerous subsequent works.

Here we use the MFT to establish the form of the Boltzmann-Gibbs distribution for the DCG at equilibrium.  We also apply the MFT to the calculation of the variance of nonequilibrium fluctuations of the excess number of particles on the positive semi-axis when starting from a (either deterministic, or random) constant density at $t=0$. Finally,  we determine the probability distribution of, and the most probable density history leading to, a \emph{large deviation} in the form of  \emph{void} of a given size in an initially uniform DCG at equilibrium.

Here is a plan of the remainder of this paper. In Sec. \ref{deterministic} we
present some similarity solutions of the deterministic Eqs.~(\ref{MF})  and (\ref{Dipole-d}), which involve infinite and finite mass, and discuss their properties. Sections \ref{MFT} and  \ref{MFTeq} deal with fluctuations in the DCG. Starting  from fluctuational hydrodynamics, as described by the Langevin equation~(\ref{Langevin0}), we introduce in Sec. \ref{MFT}  the problem of full statistics of the excess number of particles on the positive semi-axis. Using this setting as an example, we formulate the MFT of large deviations in the DCG and calculate the variance of the excess number of particles. Section \ref{MFTeq} is devoted to the MFT at equilibrium. Here we introduce the free energy density of the DCG at equilibrium  and determine the probability distribution of, and the most probable density history leading to, the formation of a void in an initially uniform gas. Section \ref{summary} presents a brief summary and discussion of our main results. Some technical details
of the derivation of the MFT equations and boundary conditions are relegated to Appendix A. In Appendix B we present an independent calculation of the variance of the particle excess directly from the Langevin equation (\ref{Langevin0}).

\section{Deterministic relaxation}
\label{deterministic}

\subsection{Infinite-mass scaling}

To start with, let us study expansion of the DCG into vacuum. Suppose that the initial gas density has the form of a step-function:
\begin{equation}\label{theta}
\rho(x,t=0) = \rho_0 \theta(-x)\,.
\end{equation}
The relaxation of this system is described by the following similarity soluton of Eq.~(\ref{MF}):
\begin{equation}\label{SS0}
\rho(x,t) = \rho_0\,R\left[\frac{x}{(\rho_0 D t)^{1/4}}\right]\,.
\end{equation}
In this case the dynamical exponent $4$ is the same as in the linear fourth-order equation $\partial_t u = -D_0 \partial_x^4 u$ originally studied in the context of surface diffusion \cite{Mullins}.

\begin{figure}[ht]
  \includegraphics[width=6.0cm]{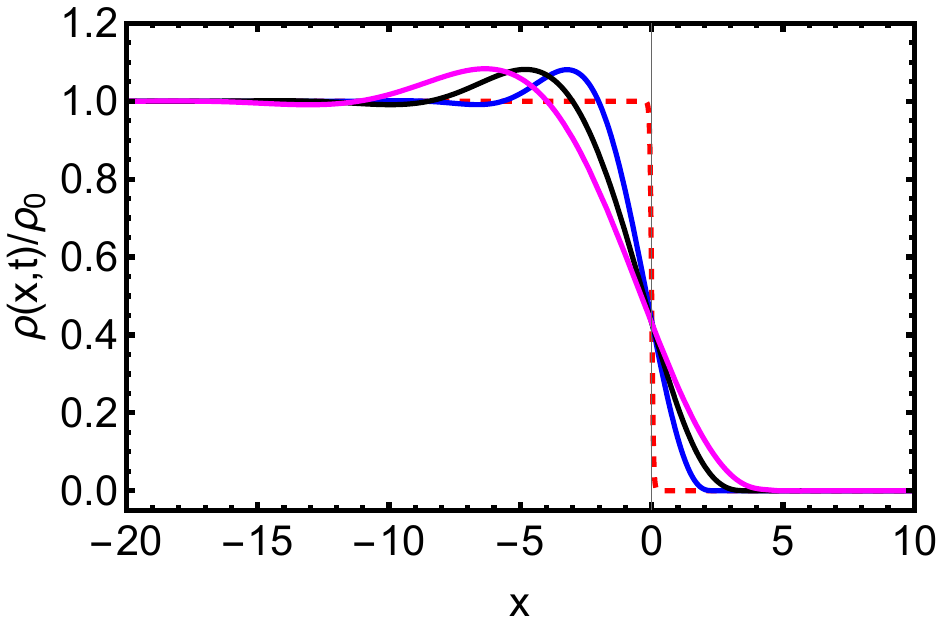}
    \includegraphics[width=6.0cm]{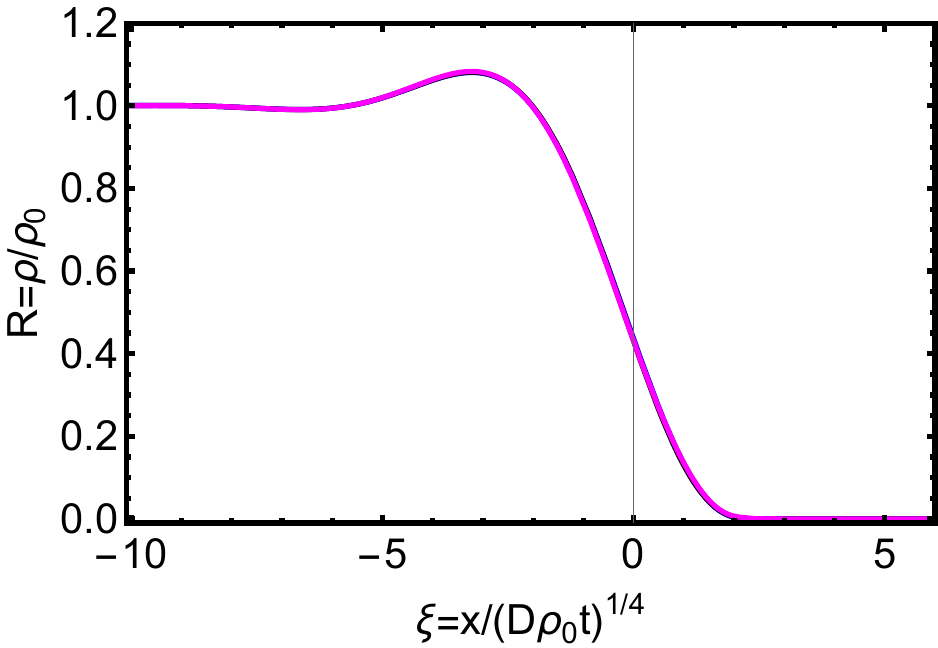}
  \caption{Top panel: Numerical solution of Eq.~(\ref{MF}) at $t=1,5$ and $15$ and $\rho(x,0)=\rho_0[1-\tanh (15\,x)]$. Bottom panel: the scaling function $R(\xi)$ obtained by plotting the same three profiles against the similarity coordinate $\xi$.}
  \label{Rtheta}
\end{figure}

The dimensionless scaling function $R(\xi)$ obeys an ordinary differential equation (ODE),
\begin{equation}\label{ODE0}
\frac{1}{4}\xi R'(\xi)= \frac{d^2}{d\xi^2}\left[R R''-(R')^2\right]\,.
\end{equation}
The boundary conditions are $R(-\infty) = 1$, $R'(-\infty) = 0$ and $R(+\infty) = R'(+\infty) = 0$. The scaling function $R(\xi)$ can be obtained by solving Eqs. (\ref{ODE0}) with these boundary conditions numerically. Alternatively, we can solve numerically the full time-dependent PDE (\ref{MF}) after bringing it to a dimensionless form by rescaling $\rho_0 x \to x$, $\rho_0^5 Dt \to t$ and $\rho/\rho_0 \to \rho$.  Figure \ref{Rtheta} gives an example of such a time-dependent solution for the rescaled initial condition $\rho(x,t=0)=1-\tanh (15\,x)$. The top panel shows this initial condition and the resulting density profiles at rescaled times $t=5$, $10$ and $15$. The bottom panel shows the same three density profiles, but plotted against the similarity coordinate $\xi$. As one can see, the profiles collapse into a single curve, which describes the scaling function $R(\xi)$. Salient features of this similarity solution are its oscillatory decay at $x\to -\infty$ and its semi-compact support: the solution is defined for $-\infty<\xi<\xi_* \simeq 2.5$. The asymptotic of the solution near the edge is $R(\xi) \simeq (\xi_*/48)  (\xi_*-\xi)^3$, so that the first and second derivatives of $R$ vanish at $\xi=\xi_*$ alongside with $R$.

\subsection{Finite-mass scaling}
\label{Nparticles}

The long-time evolution of a system with a finite number $N$ of particles is described by the similarity solution
\begin{equation}\label{SS1}
\rho(x,t) = \frac{N^{4/5}}{(Dt)^{1/5}}\,R\left[\frac{x}{(NDt)^{1/5}}\right]\,,
\end{equation}
which exhibits a different dynamical exponent $5$ \cite{HLR2024}.  The dimensionless scaling function $R(\xi)$ obeys the normalization condition
\begin{equation}\label{normalization}
\int_{-\infty}^{\infty} R(\xi)\,d\xi = 1\,,
\end{equation}
following from the conservation of the total number of particles.
The form (\ref{SS1}) is exact if the system starts from the initial condition
\begin{equation}\label{delta}
\rho(x,t=0) = N \delta(x)\,.
\end{equation}
Otherwise, Eq.~(\ref{SS1}) describes a long-time asymptotic of the solution \cite{Barenblatt}.
For the scaling function $R(\xi)$ we obtain an ODE
\begin{equation}\label{ODE1}
\frac{1}{5}\frac{d}{d\xi} \left(\xi\,R\right)= \frac{d^2}{d\xi^2}\left[R R''-(R')^2\right]\,.
\end{equation}
Integrating once, we obtain
\begin{equation}\label{ODE1a}
\frac{1}{5} \xi\,R = \frac{d}{d\xi}\left[R R''-(R')^2\right] = R R''' -R'R''\,,
\end{equation}
while the integration constant must be zero.  Since $R(\xi)$ is an even function, we can solve Eq.~(\ref{ODE1a}) on the half-line $\xi>0$
with the boundary conditions
\begin{equation}\label{BC1}
R(0) = a>0\,,\quad R'(0)=0\,,\quad R(\xi\to \infty)=0\,.
\end{equation}
The solution must be nonnegative, and the a priori unknown constant $a$ is to be determined from the normalization condition~(\ref{normalization}). The nonnegativity of the solution and the boundary condition $R(\xi\to \infty)=0$ demand that the solution have a compact support. At the edges of support both $R(\xi)$ and $R'(\xi)$ must vanish, thus providing continuity of the flux, see Eq.~(\ref{ODE1a}).

\begin{figure}[ht]
  \includegraphics[width=6.0cm]{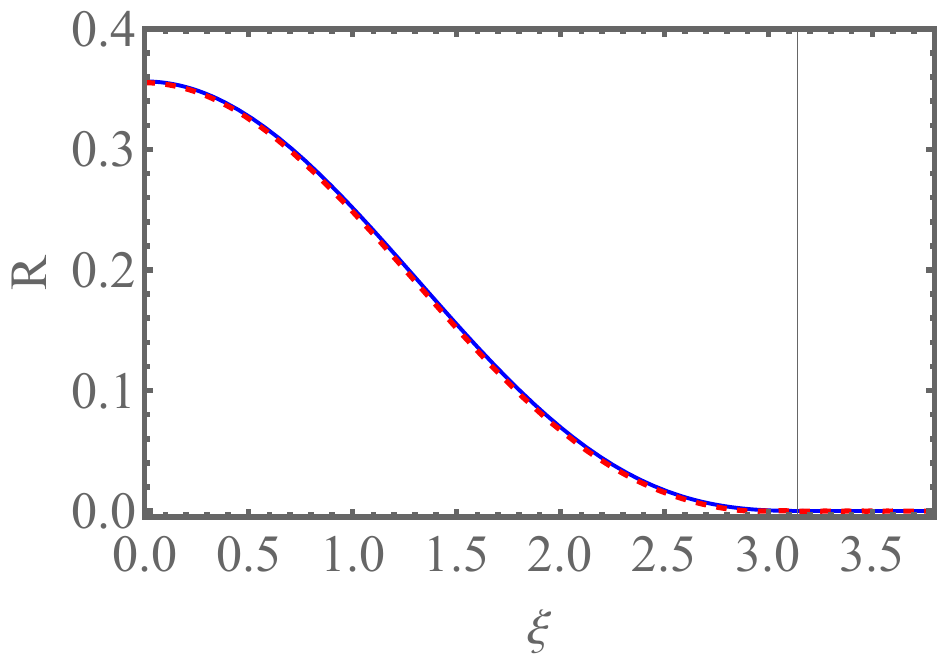}
  \caption{The scaling function $R(\xi)$ of the similarity solution~(\ref{SS1}) found by numerically solving the problem (\ref{ODE1a}) and (\ref{BC1}) (solid line) and the full time-dependent PDE (\ref{MF}) with a localized initial condition (dashed line).  Only the positive-$\xi$ region is shown.}
  \label{Rdelta}
\end{figure}

As one can check, the ODE~(\ref{ODE1a}) remains invariant under rescaling $\xi \to C^{-1/4} \xi$ and $R \to C^{-1} R$, where $C>0$. Therefore, once we have found  the solution $R_1(\xi)$  of the problem (\ref{ODE1a}) and (\ref{BC1}) for $a=1$, we can find the solution $R_a(\xi)$ for arbitrary $a$ by the rescaling transformation
\begin{equation}\label{Ra}
R_a(\xi) = a R_1\left(\frac{\xi}{a^{1/4}}\right)\,.
\end{equation}
Using Eqs.~(\ref{normalization}) and (\ref{Ra}), we obtain
\begin{equation}\label{a}
a=\left[2\int_0^{\infty} R_1(\xi)\,d\xi \right]^{-4/5}\,,
\end{equation}
so what remains is to find $R_1(\xi)$. This can be achieved numerically by the shooting method. We set $a=1$ and trade the boundary condition at infinity $R_1(\infty)=0$ for the condition $R''(0) = \gamma$, where  $\gamma<0$ serves as the shooting parameter. Having found $R_1(\xi)$ and employing Eqs.~(\ref{Ra}) and (\ref{a}), we obtained the numerical solution shown in  Fig. \ref{Rdelta}. Here $a=0.356\dots $, while the edges of support are at $|\xi|=\xi_* = 3.140\dots $. (The proximity of the latter number to $\pi$ raises curiosity but most likely coincidental.)
The asymptotic of $R(\xi)$ near the edges is $R(\xi) \simeq (\xi_*/60)  (\xi_*-|\xi|)^3$, so that the first and second derivatives of $R$ vanish alongside with $R$ at $|\xi|=\xi_*$ similarly to the step-like solution of the previous subsection.

\section{Macroscopic fluctuation theory. Fluctuations of excess number of particles}
\label{MFT}

Fluctuation hydrodynamics of the DCG is described by the Langevin equation which has been recently derived in Ref.~\cite{HLR2024}:
\begin{equation}\label{Langevin0}
\partial_{t}\rho=-D\partial_x^2\left[\rho \partial_x^2 \rho-(\partial_x \rho)^2\right]+ \sqrt{2D}\,\partial_x^2\left[\rho \eta(x,t)\right]\,,
\end{equation}
where $\eta(x,t)$ is a white Gaussian noise, $\langle\eta(x_1,t_1)\eta(x_2,t_2)\rangle = \delta(x_1-x_2) \delta(t_1-t_2)$, and we confine ourselves to one spatial dimension.

As we will show in Sec. \ref{MFTeq}, the equilibrium state of the DCG  can be described by the Boltzmann-Gibbs distribution with a well-defined free-energy density $F(\rho)$.  That is, when the gas is in equilibrium at density $\rho_0$, the probability density of observing an arbitrary
density profile $\rho_*(x)$ is given by $-\ln {\mathcal P}[\rho_*(x)] \simeq S_{\text{eq}}$, where
\begin{equation}\label{BG}
    S_{\text{eq}} \!\simeq \!\int_{-\infty}^{\infty} dx \left[F(\rho_*(x))-F(\rho_0)-F^{\prime}(\rho_0)(\rho_*(x)-\rho_0)\right].
\end{equation}
Remarkably, the free energy density of this gas,
\begin{equation}\label{F}
F(\rho) = \rho \ln \rho -\rho\,,
\end{equation}
coincides with that of the lattice gas of noninteracting random walkers (RWs). The latter is described by the more familiar second-order Langevin equation \cite{Spohn91}
\begin{equation}\label{LangevinRW}
\partial_{t}\rho= \partial_{x}\left[D\partial_x\rho + \sqrt{2D\rho}\, \eta(x,t)\right]\,.
\end{equation}
Plugging Eq.~(\ref{F}) for $F(\rho)$ into Eq.~(\ref{BG}), we obtain
\begin{equation}\label{Seq1}
S_{\text{eq}} = \int_{-\infty}^{\infty} dx \left[\rho_*(x) \ln \frac{\rho_*(x)}{\rho_0}+\rho_0-\rho_*(x)\right]\,,
\end{equation}
which coincides with the corresponding expression for the gas of noninteracting RWs.

Large deviations in the DCG can be described by the MFT \cite{bertini}. The MFT is a weak-noise theory which is based on the Langevin equation (\ref{Langevin0}) and relies on a problem-specific small parameter, that we identify below. When such a parameter is present, the probability distribution of interest can be approximately determined via a saddle-point evaluation of the exact path integral corresponding to Eq. (\ref{Langevin0}) and the problem-specific constraints.

We will introduce the MFT for the DCG on the example of the statistics of the excess number of particles $K$ on the positive semi-axis at specified time $T$. One can consider two different settings. In the first of them we start from a constant gas density $\rho_0$ and condition the process on the value of the integral
\begin{equation}\label{constraint0}
\int_0^{\infty}\left[\rho(x,T) - \rho_0\right] \,dx = K.
\end{equation}
at time $T$. This setting corresponds to infinite total mass of the system.

In the second setting (a finite total mass) there is a large but finite number of particles, $N\gg 1$, in the system. Here the particle excess condition at $t=T$ is
\begin{equation}\label{constraintfinite}
\int_0^{\infty} \rho(x,T)\,dx-\frac{N}{2}=K\,,
\end{equation}
where $-N/2\leq K\leq N/2$.

As we will see shortly, the scaling behavior of the particle excess statistics in the two settings is quite different. In the infinite-mass setting the result also depends on whether the initial condition is quenched: that is, deterministically prepared, or annealed: that is, randomly sampled from the equilibrium distribution of the gas at density $\rho_0$.

\subsection{Infinite-mass scaling}

Let us rescale the variables: $t/T \to t$, $x/(\rho_0 DT)^{1/4} \to x$, and $\rho/\rho_0 \to \rho$. In the new variables Eqs.~(\ref{Langevin0}) and~(\ref{constraint0}) become
\begin{equation}\label{Langevin1}
\partial_{t}\rho=-\partial_{x}^{2}\left(\rho^2 \partial_x^2 \ln \rho\right)+ \sqrt{2}\,\left(\rho_0^5 D T\right)^{-1/8}\partial_x^2\left[\rho \eta(x,t)\right]
\end{equation}
and
\begin{equation}\label{constraint1}
\int_0^{\infty}\left[\rho(x,1) - 1\right] \,dx = j \equiv \frac{K}{(\rho_0^5 DT)^{1/4}} ,
\end{equation}
respectively.  At long time the noise becomes effectively weak. The presence of a large parameter $(\rho_0^5 D T)^{1/4}\gg 1$, which is a typical number of particles within a region with the length scale $(\rho_0 DT)^{1/4}$, makes it possible to develop the MFT, that is to perform a saddle-point evaluation of the exact path integral for Eq.~(\ref{Langevin1}) with account of the constraint~(\ref{constraint1}). The calculation boils down to minimization of the action functional (see Appendix A for details) and leads to the following Hamilton's equations that describe the optimal (that is, the most likely) path of the system conditioned on Eq.~(\ref{constraint1}):
\begin{eqnarray}
\partial_t \rho &=& \partial_x^2 \left[-\rho\partial_x^2 \rho + (\partial_x \rho)^2+2\rho^2 v\right]\,,\label{eqrho}\\
\partial_t v &=& \partial_x^2 \left[v\partial_x^2\rho +\partial_x^2(v\rho)
+2\partial_x(v \partial_x\rho)-2v^2\rho\right]\,.\label{eqv}
\end{eqnarray}
The boundary condition at the observation time $t=1$ is
\begin{equation}\label{T}
v(x,1) = \lambda \,\delta'(x)\,,
\end{equation}
where $\lambda$ is a Lagrange multiplier, introduced when accommodating the constraint (\ref{constraint1}),
and $\delta'(x)$ is the $x$-derivative of the delta function. The quenched initial condition is
\begin{equation}\label{0}
\rho(x,0) = 1\,.
\end{equation}
The annealed initial condition is introduced shortly.

Once the optimal path is found, the probability distribution is given, up to a pre-exponent,
by the action along the optimal path: $ -\ln  \mathcal{P} (K,T,\rho_0) \simeq S$, where
\begin{equation}
S=  \left(\rho_0^5 D T\right)^{1/4} s(j)\,,
  \label{action1}
\end{equation}
where
\begin{equation}\label{s1}
s(j) = \int_0^1 dt \int_{-\infty}^{\infty} dx \, \rho^2(x,t) v^2(x,t)\,.
\end{equation}
The rescaled excess of the number of particles $j$ is defined in Eq.~(\ref{constraint1}).

For the annealed initial condition with \emph{average} rescaled density $1$ the full action includes the cost of creating the optimal initial condition $\rho(x,0)$: $S_{\text{annealed}}=S+S_0$, where $S$ is the dynamical action, described by Eqs.~(\ref{action1}) and (\ref{s1}), and
\begin{equation}\label{S0}
S_0=\int_{-\infty}^{\infty} dx \left\{F[\rho(x,0)]-F(1)-F^{\prime}(1)[\rho(x,0)-1]\right\},
\end{equation}
As a result, Eq.~(\ref{0}) gives way to a different condition \cite{DG2009b}, which describes a relation between the a priori unknown $\rho(x,0)$ and $v(x,0)$:
\begin{equation}\label{0annealed}
v(x,0) -\frac{d^2}{dx^2}\ln \rho(x,0)=\lambda \delta'(x)\,.
\end{equation}

\subsection{Finite-mass scaling}

In this case the parameter $\rho_0$ is absent, and the rescaling of variables
is different:
\begin{equation}\label{finiterescaling}
\frac{t}{T} \to t, \quad \frac{x}{(NDT)^{1/5}} \to x \quad \text{and}\quad \frac{(DT)^{1/5}\rho}{N^{4/5}}\to \rho\,.
\end{equation}
Notice that the infinite-mass dynamical exponent $4$ gives way to the finite-mass exponent $5$. 
The rescaled Eqs.~(\ref{Langevin0}) and (\ref{constraintfinite}) become
\begin{equation}\label{Langevin2}
\partial_{t}\rho=-\partial_{x}^{2}\left(\rho^2 \partial_x^2 \ln \rho\right)+ \sqrt{\frac{2}{N}}\, \partial_x^2\left[\rho \eta(x,t)\right]\,,
\end{equation}
and
\begin{equation}\label{constraint2}
\int_0^{\infty}\rho(x,1)\,dx -\frac{1}{2} = \frac{K}{N} \equiv j\,,
\end{equation}
where $|j|\leq 1/2$. Here the saddle-point expansion, leading to the MFT, relies on the large parameter $N\gg 1$.

The rescaled MFT equations coincide with Eqs.~(\ref{eqrho}) and (\ref{eqv}). The boundary condition (\ref{T}) is also the same, but Eq.~(\ref{0}) gives way to the condition
\begin{equation}\label{0finite}
\rho(x,0) = \delta(x)\,.
\end{equation}
Here the probability distribution of the number of particles, transferred to the right, is independent of the observation time $T$: $ -\ln  \mathcal{P} (K,N) \simeq S$, where
\begin{equation}
S= N s(j)\,,
  \label{action2}
\end{equation}
and $s(j)$ is again described by Eq.~(\ref{s1}).

\subsection{Variance of particle excess}
\label{typicalK}

The statistics of typical (that is small) fluctuations of the excess number of particles $K$ is Gaussian, and its variance
scales as a characteristic number of particles involved. In the infinite-mass case this is the typical
number of particles over the dynamical length scale $(\rho_0 DT)^{1/4}$, that is $\text{var}_K \sim (\rho_0^{5}DT)^{1/4}$. In the finite-mass case  it is simply $\text{var}_K \sim N$. Essentially, the role of theory  (a first-order perturbation theory in $|\lambda|\ll 1$, developed in Ref. \cite{KrMe}) is to
provide the numerical coefficients $O(1)$ in these expressions.

For the infinite-mass case the calculations are straightforward. Upon linearization with respect to $\lambda$, the MFT equations (\ref{eqrho}) and (\ref{eqv}) become
\begin{eqnarray}
\partial_t \rho &=& -\rho_0\partial_x^4 \delta \rho +\rho_0^2 \partial_x^2 v\,,\label{eqrholin}\\
\partial_t v &=& \partial_x^4 v\,.\label{eqvlin}
\end{eqnarray}
We can solve Eq.~(\ref{eqvlin}) backward in time with the ``initial condition" (\ref{T}). The solution can be obtained by differentiating with respect to $x$ the previously known solution for the initial condition in the form of a delta-function (see e.g. Ref. \cite{MV2016} and Eq. (\ref{G}) in Appendix B below).
The result has the similarity form
\begin{equation}\label{vss}
v(x,t) = \frac{\lambda}{(1-t)^{1/2}} \,V\left[\frac{x}{(1-t)^{1/4}}\right]\,.
\end{equation}
The scaling function $V(z)$ can be expressed via the hypergeometric function $_0F_2$:
\begin{widetext}
\begin{equation}
  V(z) = -\frac{z \Gamma \left(\frac{3}{4}\right) \,
   _0F_2\left(\frac{5}{4},\frac{3}{2};\frac{z^4}{256}\right)}{4 \pi}
   +\frac{z^3 \Gamma \left(\frac{5}{4}\right) \,
   _0F_2\left(\frac{3}{2},\frac{7}{4};\frac{z^4}{256}\right)}{24 \pi}
    -\frac{z^5 \Gamma \left(\frac{3}{4}\right) \,
   _0F_2\left(\frac{9}{4},\frac{5}{2};\frac{z^4}{256}\right)}{960 \pi}\,. \label{vlinear}
\end{equation}
\end{widetext}
Now we can evaluate the rescaled dynamical action (\ref{s1}) in terms of $\lambda$. Within the linear theory in $\lambda$,
we should replace the density $\rho(x,t)$ in Eq.~(\ref{s1}) by $1$. We obtain
\begin{eqnarray}
  s(\lambda) &=& \int_0^1 dt \int_{-\infty}^{\infty} dx \,v^2(x,t)\nonumber \\
   &=& \lambda^2 \int_0^1 \frac{dt}{(1-t)^{3/4}} \int_{-\infty}^{\infty} dz \,V^2(z)\,. \label{s1lin}
\end{eqnarray}
The integral over $t$ gives $4$. The integral over $z$ can be evaluated numerically: $\int_{-\infty}^{\infty} V^2(z) \,dz = \alpha= 0.05798\dots$. Overall, we obtain $s=4\alpha \lambda^2$.

\subsubsection{Quenched initial condition}
\label{quenched}

For the quenched setting, the formula $s=4\alpha \lambda^2$ suffices for expressing the action in terms of $j$. Indeed, using the ``shortcut relation" $ds/dj=\lambda$ (see, e.g. Ref. \cite{Vivo}),  we obtain
\begin{equation}\label{shortcut1}
\frac{ds}{d\lambda}\frac{d\lambda}{dj} = 2\alpha \lambda\,\frac{d\lambda}{dj}=\lambda\,,
\end{equation}
which gives $\lambda=j/(8 \alpha)$. As a result,
\begin{equation}\label{slinresult}
s = 4\alpha \lambda^2= \frac{j^2}{16\alpha}\,.
\end{equation}
Back to the dimensional variables, we obtain the variance of the typical fluctuations of $K$:
\begin{equation}\label{varKquenched}
\text{var} K = 8\alpha (\rho_0^5 D T)^{1/4}\,.
\end{equation}

Exactly the same result (\ref{varKquenched}) for the quenched initial condition can be obtained directly from the linearized version of the Langevin equation~(\ref{Langevin1}). This calculation is presented in Appendix \ref{appendixb}. The MFT, however, also enables one to calculate the variance for the annealed initial condition, where a direct calculation with the linearized Langevin equation does not seem to be available.

\subsubsection{Annealed initial condition}

In the annealed setting one should also take into account (the small-$\lambda$ expansions of) Eqs.~(\ref{S0}) and~(\ref{0annealed}), which are the following:
\begin{equation}\label{S0lin}
s_0=\frac{1}{2}\int_{-\infty}^{\infty} \delta \rho^2(x,t=0)\,dx
\end{equation}
and
\begin{equation}\label{0annealedlin}
v(x,t=0) -\frac{d^2}{dx^2}\delta \rho(x,t=0)=\lambda \delta'(x)\,,
\end{equation}
respectively, where $\delta \rho(x,t=0)$ is the a priori unknown small perturbation on the background of the constant density $\rho=\rho_0$, and $v(x,t=0)$ is also small.

Evaluating Eq.~(\ref{vss}) at $t=0$, plugging the result into Eq.~(\ref{0annealedlin}) and integrating the latter equation twice over $x$, we  determine the optimal initial density field $\rho(x,0)\simeq 1+\delta \rho(x,0)$, where
\begin{equation}\label{deltarho}
\delta \rho(x,0) = \lambda \left[\int_{-\infty}^x dy \int_{-\infty}^y dz \,F(z)-\theta(x)\right]\,,
\end{equation}
and $\theta(x)$ is the step-function. The double integral in this expression can be evaluated with ``Mathematica" analytically, but the result is too cumbersome to present it here. Plugging it into Eq.~(\ref{S0lin}) and evaluating the resulting integral numerically, we obtain
\begin{equation}\label{slin1}
s_0= \beta \,\lambda^2\,,
\end{equation}
where $\beta = 0.1581\dots$. The total annealed action is, therefore,
\begin{equation}\label{s1linann}
s(\lambda) = (4\alpha+\beta) \lambda^2 = \frac{j^2}{4(4\alpha+\beta)}\,,
\end{equation}
where $\alpha=0.05798\dots$, as obtained above, and we have again used the shortcut relation  $ds/dj=\lambda$ to express the action through $j$.

At given $j$, the annealed action (\ref{s1linann}) is smaller than the quenched action (\ref{slinresult}). Back in the original variables, the variance of the typical fluctuations of $K$ in the annealed case,
\begin{equation}\label{varKan}
\text{var} K = (8\alpha+2\beta) (\rho_0^5 D T)^{1/4}\,,
\end{equation}
is larger than that in the quenched case [see Eq.~(\ref{varKquenched})], as to be expected on physical grounds.

\section{MFT of large deviations at equilibrium}
\label{MFTeq}

Now we suppose that the DCG is at equilibrium and, using the MFT, evaluate
the probability density of observing a specified density profile $\rho_*(x)$. Importantly, the Hamiltonian MFT equations are still Eqs.~(\ref{eqrho}) and (\ref{eqv}) (where we return to the dimensional variables except setting $D=1$). The boundary conditions in time, however, become $\rho(x,t=-\infty) = \rho_0$ and $\rho(x,t=0) = \rho_*(x)$ \cite{KrMe,KMSvoid}.

At the microscopic level, the DCG obeys detailed balance. At the macroscopic level the detailed balance manifests itself as the Onsager-Machlup reversibility principle \cite{Kurchan}. In particular, the optimal activation path $\rho(x,t)$, leading from $\rho(x,t=-\infty) = \rho_0$ to $\rho(x,t=0)=\rho_*(x)$, must coincide with
the \emph{time-reversed} relaxation path from $\rho(x,t=-\infty)=\rho_*(x)$ to $\rho(x,t=\infty) = \rho_0$. That is, the optimal path must obey the equation
\begin{equation}\label{timereversed}
\partial_{t}\rho= \partial_{x}^{2}\left(\rho^2 \partial_x^2 \ln \rho\right)\,.
\end{equation}
Combining this equation with
Eq.~(\ref{eqrho}), we obtain the important relation
\begin{equation}\label{equilibrium1}
v(x,t) = \partial_x^2 \ln \rho(x,t)\,,
\end{equation}
which describes the equilibrium manifold of this system. The relation (\ref{equilibrium1}) brings about two important consequences. Firstly, as one can check by a direct substitution, Eq.~(\ref{eqv}) is now obeyed automatically. Secondly,
the mechanical action can be calculated as follows:
\begin{eqnarray}
  S &=& \int_{-\infty}^0 dt \int_{-\infty}^{\infty} dx \, \rho^2(x,t) v^2(x,t) \nonumber \\
  &=& \int_{-\infty}^0 dt \int_{-\infty}^{\infty} dx \, \rho^2(x,t) \left(\partial_x^2 \ln \rho\right)^2\,.
\end{eqnarray}
After two integrations by part over $x$, this expression becomes
\begin{eqnarray}
  S &=& \int_{-\infty}^0 dt \int_{-\infty}^{\infty} dx \ln \rho  \, \partial_x^2
  \left(\rho^2 \partial_x^2 \ln \rho \right) \,.
\end{eqnarray}
By virtue of Eq.~(\ref{timereversed}), this expression can be recast as
\begin{eqnarray}
  S &=& \int_{-\infty}^{\infty} dx \int_{-\infty}^0 dt \,\ln \rho\,
  \partial_t\rho \nonumber \\
 &=& \int_{-\infty}^{\infty} dx \int_{-\infty}^0 dt \,  \partial_t \left(\rho\ln\rho -\rho\right)\,.
\end{eqnarray}
Performing the integration over time, we arrive at the announced Boltzmann-Gibbs relation (\ref{Seq1}).

\begin{figure}[ht]
  \includegraphics[width=6.0cm]{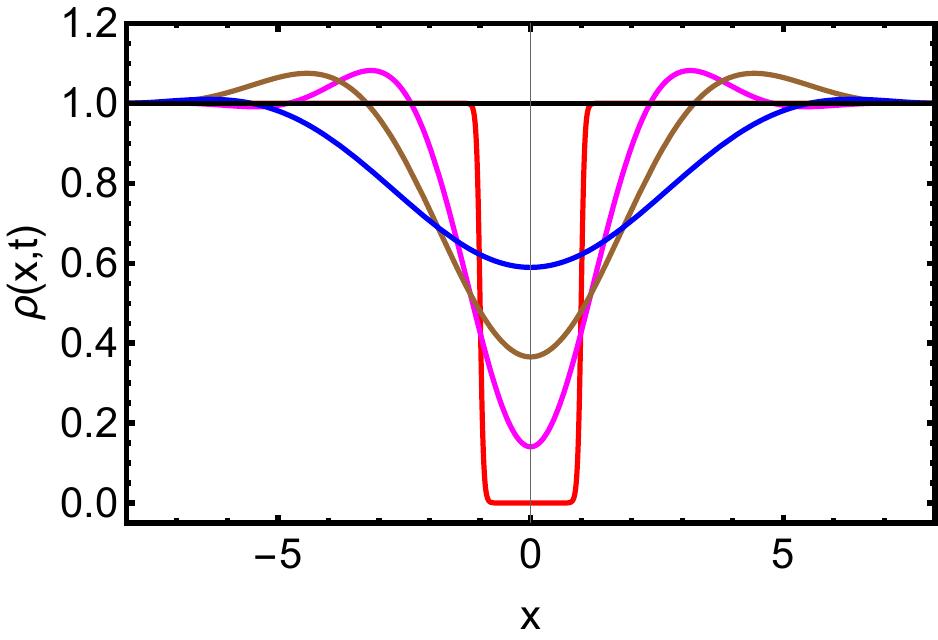}
  \caption{The optimal path of void formation at $t=0$ and $\rho_0=L=D=1$. Shown is the optimal density profile at times $t=-\infty$ (black), $-6$ (blue), $-2$ (brown), $-1/5$ (magenta) and $0$ (red).}
  \label{voidfig}
\end{figure}

As a simple but instructive example, let us evaluate the probability distribution $\mathcal{P}_{\text{void}}$ of observing a void of size $2L$ in a uniform gas at equilibrium:
\begin{equation}
\label{void}
\rho_*(x) =
\begin{cases}
0,   & |x|<L ,\\
\rho_0 ,  & |x|>L.
\end{cases}
\end{equation}
Using Eq.~(\ref{Seq1}), we obtain $\mathcal{P}_{\text{void}}\sim \exp(-2\rho_0 L)$, as in the gas of noninteracting RWs \cite{KMSvoid}.   The optimal path of the system toward the void formation, $\rho(x,t)$ is obtained by the time reversal of the relaxation dynamics of the void, see Eq.~(\ref{timereversed}). We computed this optimal path numerically, and the results are shown in Fig. \ref{voidfig}. Noticeable are spatial oscillations of the density, which are absent in the optimal path of the void formation in the gas of noninteracting RWs, see Ref. \cite{KMSvoid}.

\section{Summary and Discussion}
\label{summary}

We have considered some basic macroscopic relaxation and fluctuation properties of the mass- and dipole-conserving stochastic lattice gas. We have extended the MFT approach to this system. Using some carefully selected examples, we have demonstrated how one can apply the MFT both to typical fluctuations of the DPG, and to its large deviations. We hope that the MFT formalism will find additional applications to particular  settings relevant to experiments.

We would like to conclude this work with a fascinating observation. Let us briefly recall one basic property of diffusive lattice gases that conserve only the number of particles. The fluctuational hydrodynamics of such gases is described by the Langevin equation \cite{Spohn91}
\begin{equation}\label{Langevinmassonly}
\partial_{t}\rho= \partial_{x}\left[\mathcal{D}(\rho)\partial_x\rho + \sqrt{\sigma(\rho)}\, \eta(x,t)\right]\,,
\end{equation}
which generalizes Eq.~(\ref{LangevinRW}) to a broad class of diffusive lattice gases.
The free energy density $F(\rho)$ of this class of gases is determined by the diffusion coefficient $\mathcal{D}(\rho)$ and the mobility $\sigma(\rho)$ via the Einstein relation \cite{Spohn91}
\begin{equation}\label{einstein}
F''(\rho) = \frac{2\mathcal{D}(\rho)}{\sigma(\rho)}\,.
\end{equation}
In the particular case of noninteracting RWs, see Eq.~(\ref{LangevinRW}), one has $\mathcal{D}(\rho)= D =\text{const}$, and $\sigma(\rho) = 2 D \rho$. Then Eq.~(\ref{einstein}) yields the free energy density $F(\rho) = \rho \ln \rho -\rho$.  The remarkable coincidence of this expression with the free energy density~(\ref{F}) for the DCG  may suggest the validity of the Einstein relation (\ref{einstein}) for dipole-conserving gases at the level of the Langevin equation~(\ref{Langevin1}). Indeed, if, by analogy with Eq.~(\ref{Langevinmassonly}), we identify the functions $\mathcal{D}(\rho) = D \rho$ (which enters the highest-derivative term) and $\sigma(\rho)=2 D \rho^2$, then Eq.~(\ref{einstein}) correctly reproduces the free energy density (\ref{F}). This observation hints at possible additional surprises in the studies of the dipole-conserving lattice gases.

\section*{Acknowledgments}

The author is very grateful to P. L. Krapivsky for stimulating discussions and comments. This research was supported by the Israel Science Foundation (Grant No. 1499/20).

\vspace{0.5 cm}

\appendix

\begin{widetext}

\section{Derivation of Eqs.~(\ref{eqrho}) and (\ref{eqv}) and boundary conditions}
\label{appendix}
\noindent Our starting point is Eq.~(\ref{Langevin1}). One way to deal with the second derivative in the noise term is
to define the ``second-order potential" $\psi(x,t)$ by the relation $\rho(x,t) = \partial_x^2 \psi(x,t)$
and rewrite Eq.~(\ref{Langevin1}) in terms of $\psi(x,t)$:
\begin{equation}\label{Langevinpsi}
\partial_t\psi=-\partial_{x}^{2}\psi \,\partial_x^4\psi+ (\partial_x^3 \psi)^2+\sqrt{2 \epsilon} \,\partial_x^2 \psi \,\eta(x,t)\,,
\end{equation}
where $\epsilon=\left(\rho_0^5 D T\right)^{-1/4}\ll 1$. The probability density of a realization of the (dimensionless) Gaussian white noise $\eta(x,t)$ is
\be
\mathcal{P}\left[\eta\right]\sim\exp\left(-\int_{0}^{1}dt\int_{-\infty}^{\infty}dx\frac{\eta^{2}}{2}\right).
\ee
Expressing $\xi$ through $\psi$ and its derivatives from Eq.~(\ref{Langevinpsi}), we obtain
\be
\mathcal{P}[\psi(x,t)] \sim \exp\left\{-\frac{1}{4\epsilon}
\int_{0}^{1}dt\int_{-\infty}^{\infty}dx
\frac{\left[\partial_t\psi+\partial_{x}^{2}\psi \,\partial_x^4\psi- (\partial_x^3 \psi)^2\right]^2}{(\partial_x^2 \psi)^2}\right\}\,.
\label{functional}
\ee
Employing the small parameter $\epsilon \ll 1$, we can apply the saddle-point approximation. The calculations boil down to a minimization of the action functional~
\begin{equation}
 s_{\lambda}[\psi(x,t)] = \frac{1}{4}
\int_{0}^{1} dt \int_{-\infty}^{\infty} dx
\frac{\left[\partial_t\psi+\partial_{x}^{2}\psi \,\partial_x^4\psi- (\partial_x^3 \psi)^2\right]^2}{(\partial_x^2 \psi)^2}
-\lambda\int_{-\infty}^{\infty} \theta(x) \left[\partial_x^2 \psi(x,1)-\partial_x^2 \psi(x,0)\right]\,dx\,,
\label{functional1}
\end{equation}
where the constraint (\ref{constraint1}) is incorporated with the help of a Lagrange multiplier $\lambda$.
Let us introduce the second derivative $\partial_x^2 p$ of the canonical momentum density $p(x,t)$ by differentiating the Lagrangian of the action functional (\ref{functional1}) with respect to $\psi_t$ \cite{LL}. We obtain
\be
\label{pdef}
\partial_{x}^2 p=\frac{\partial_t\psi+\partial_{x}^{2}\psi \,\partial_x^4\psi- (\partial_x^3 \psi)^2}
{2(\partial_x^2 \psi)^2}\,.
\ee
Denoting $\partial_x^2 p(x,t) \equiv v(x,t)$ and going back from $\psi(x,t)$ to the density $\rho(x,t)$, we obtain Eq.~(\ref{eqrho}).

The Hamiltonian $H$, corresponding to the  first term of the action functional (\ref{functional1}), is the following \cite{LL}:
\be
H=\int_{-\infty}^{\infty} dx \left[v^2 \partial_x^2\psi -v \partial_{x}^{2}\psi \,\partial_x^4\psi+v (\partial_x^3 \psi)^2 \right]\,.
\ee
Equation~(\ref{eqv}) can be obtained by taking the minus variational derivative of this Hamiltonian with respect to $\psi$, and then going back from $\psi$ to $\rho$.

This derivation, however, ignores possible boundary terms at $t=0$ and $t=1$, arising when calculating the linear variation of the constrained action $s_{\lambda}[\psi(x,t)]$. One boundary term at $t=1$ comes from the term proportional to $\lambda$ in Eq.~(\ref{functional1}). An additional term at $t=1$ results from integration by parts of the term that includes $\partial_t\delta \psi(x,t)$, where $\delta \psi(x,t)$ is the linear variation of $\psi(x,t)$.  These two terms yield the boundary condition~(\ref{T}).

For the quenched initial condition~(\ref{0}) there are no boundary terms at $t=0$.
In the annealed setting there are two boundary terms at $t=0$. One of them arises from integration by parts of the term that includes $\partial_t\delta \psi(x,t)$, while the other comes from the variation of the term proportional to $\lambda$ in Eq.~(\ref{functional1}). (Indeed, the initial optimal density $\rho(x,t=0)$ in this case is a priori unknown.) These two terms lead to the annealed initial condition~(\ref{0annealed}).

Finally, the action~(\ref{s1}) is given
by the first term in Eq.~(\ref{functional1}), rewritten in terms of $\rho$ and $v$.

\section{Variance of particle excess number from the Langevin equation}
\label{appendixb}

Typical, small fluctuations of $K$ are caused by typical, small fluctuations of the density profile $\rho(x,t)$. In order to account for such fluctuation,  one can linearize the Langevin equation~(\ref{Langevin1}) around the equilibrium state $\rho=1$: $\rho(x,t) = 1+u(x,t)$, where $|u|\ll 1$ \cite{KrMe}. The linearized equation is
\begin{equation}\label{Langevin1a}
\partial_{t} u(x,t)=-\partial_{x}^{4}u(x,t)+ \sqrt{2\epsilon}\,\partial_x^2\eta(x,t)\,,
\end{equation}
where $\epsilon = \left(\rho_0^5 DT\right)^{-1/4} \ll 1$. The rescaled particle excess number  at $t=1$, see Eq.~(\ref{constraint1}), becomes
\begin{equation}\label{constraint1a}
j \equiv \epsilon K = \int_0^{\infty}u(x,1) \,dx\,.
\end{equation}
Using the new variable,
\begin{equation}\label{psia}
\psi(x,t) = \int_{-\infty}^x dy\int_{-\infty}^y dz\,u(z,t)\,,
\end{equation}
we can rewrite Eq.~(\ref{Langevin1a}) as
\begin{equation}\label{Langevin1b}
\partial_{t} \psi(x,t)=-\partial_{x}^{4}\psi(x,t)+ \sqrt{2\epsilon}\,\eta(x,t)\,,
\end{equation}
and formally solve it for a given realization of noise $\eta(x,t)$ and the quenched initial condition $\psi(x,t=0)=0$.  In particular, at rescaled time $t=1$ we obtain
\begin{equation}\label{solpsi}
\psi(x,t=1) = \sqrt{2\epsilon} \int_0^1 dt \int_{-\infty}^{\infty} dy\,\eta(y,t) G(x-y,1-t)\,,
\end{equation}
where $G(z,\tau)$ is the Green's function of the homogeneous equation, corresponding to Eq.~(\ref{Langevin1a}).
That is, $G(z,\tau)$ solves the equation $\partial_{\tau} G(z,\tau)=-\partial_{z}^{4}G(z,\tau)$ with the
initial condition $G(z,\tau=0)=\delta(z)$. The explicit form of $G(z,\tau)$ is the following:
\begin{equation}\label{G}
G(z,\tau)=\frac{\Gamma \left(\frac{5}{4}\right) \,
   _0F_2\left(\frac{1}{2},\frac{3}{4};\frac{z^4}{256 \tau
   }\right)}{\pi \tau^{1/4}}-\frac{z^2 \Gamma \left(\frac{3}{4}\right)
   \, _0F_2\left(\frac{5}{4},\frac{3}{2};\frac{z^4}{256 \tau }\right)}{8 \pi
   \tau ^{3/4}}\,.
\end{equation}
Plugging Eq.~(\ref{solpsi}) into the constraint~(\ref{constraint1a}) and performing the integration over $x$, we obtain the fluctuating particle excess number
at $t=1$: $j=-\partial_x \psi(x=0,t=1)$. The variance of these fluctuations is
\begin{equation}\label{vara}
 \text{Var}_K = \langle \left[\partial_x \psi(0,1)\right]^2\rangle
  = 2\epsilon\int_0^1 dt_1 \int_0^1 dt_2 \int_{-\infty}^{\infty}dy_1 \int_{-\infty}^{\infty}dy_2
  \left\langle \eta(y_1,t_1) \eta(y_2,t_2) \right\rangle \frac{\partial G(y_1,1-t_1)}{\partial y_1}\times \frac{\partial G(y_2,1-t_2)}{\partial y_2}\,.
\end{equation}
Since $ \left\langle \eta(y_1,t_1) \eta(y_2,t_2) \right\rangle = \delta(y_1-y_2) \delta(t_1-t_2)$, Eq.~(\ref{vara}) simplifies to
\begin{equation}\label{varb}
 \text{Var}_K = 2 \epsilon \int_0^1 dt \int_{-\infty}^{\infty} dx \left[\frac{\partial G(x,1-t)}{\partial x}\right]^2\,.
\end{equation}
Up to a constant factor, this double integral is exactly the same as in the linearized MFT calculation, see Eq.~(\ref{s1lin}). Equation~(\ref{varb}) gives $\text{Var}_K=8 \alpha \epsilon$ (where $\alpha =0.05798\dots$), which perfectly coincides with our MFT result~(\ref{varKquenched}).

\end{widetext}


\begin{thebibliography}{99}
\bibitem{jamming} For a coarse-grained description to apply the particle density must be sufficiently large
so as to avoid jamming, see e.g. Ref. \cite{Arita}.
\end{thebibliography}
\end{document}